# *Top-Down* Multilevel Simulation of Tumor Response to Treatment in the Context of *In Silico* Oncology

**Georgios Stamatakos** [1, §]

[1, §] *In Silico* Oncology Group, Laboratory of Microwaves and Fiber Optics, Institute of Communication and Computer Systems, School of Electrical and Computer Engineering, National Technical University of Athens. Iroon Polytechniou 9, Zografos GR 157 80, Greece,

[§]**Corresponding Author:**
Georgios S. Stamatakos, PhD
*In Silico* Oncology Group
Laboratory of Microwaves and Fiber Optics
Institute of Communication and Computer Systems
School of Electrical and Computer Engineering
National Technical University of Athens
Iroon Polytechniou 9, Zografos GR 157 80
Greece
Tel: (+30) 210 772 2288
Fax: (+30) 210 772 3557
E-Mail: gestam@central.ntua.gr
URL: www.in-silico-oncology.iccs.ntua.gr

**Abbreviations:** *A* = apoptosis; *ACGT* = Advancing Clinico-Genomic Trials on Cancer; *AHF* = Accelerated Hyper-Fractionation; *BIR* = British Institute of Radiology; *Cpav* = Average Plasma Concentration; *CT* = Computerized Tomography; *EC* = European Commission, *GBM* = Glioblastoma Multiforme; $G_0$ or *G0* = dormant phase ; $G_1$ or *G1* = Gap 1 phase of the cell cycle; $G_2$ or *G2* = Gap 2 phase; *HF* = Hyper-Fractionation, *IEEE* = Institute of Electrical and Electronics Engineers; *M* = Mitosis; *MRI* = Magnetic Resonance Imaging; *N* = necrosis; *PET* = Positron Emission Tomography; *R&D* = Research and Development; *RTOG* = Radiation Therapy Oncology Group; *S* = DNA synthesis phase; *TDS* = Time Delay in the S phase compartment; *TMZ* = Temozolomide

Dedicated to Werner Duechting for his inspiring work on tumor modeling and his inspiration inducing character on the occasion of his 75th birthday

**ABSTRACT**

The aim of this chapter is to provide a brief introduction into the basics of a *top-down* multilevel tumor dynamics modeling method primarily based on discrete entity consideration and manipulation. The method is clinically oriented, one of its major goals being to support patient individualized treatment optimization through experimentation *in silico* (=on the computer). Therefore, modeling of the treatment response of clinical tumors lies at the epicenter of the approach. Macroscopic data, including i.a. anatomic and metabolic tomographic images of the tumor, provide the framework for the integration of data and mechanisms pertaining to lower and lower biocomplexity levels such as clinically approved cellular and molecular biomarkers. The method also provides a powerful framework for the investigation of multilevel (multiscale) tumor biology in the generic investigational context. The *Oncosimulator*, a multiscale physics and biomedical engineering concept and construct tightly associated with the method and currently undergoing clinical adaptation, optimization and validation, is also sketched. A brief outline of the approach is provided in natural language. Two specific models of tumor response to chemotherapeutic and radiotherapeutic schemes are briefly outlined and indicative results are presented in order to exemplify the application potential of the method. The chapter concludes with a discussion of several important aspects of the method including i.a. numerical analysis aspects, technological issues, model extensions and validation within the framework of actual running clinico-genomic trials. Future perspectives and challenges are also addressed.

## 1. INTRODUCTION

The majority of cancer modeling techniques developed up to now adopt the straightforward *bottom-up* approach focusing on the better understanding and quantification of rather microscopic tumor dynamics mechanisms and the investigation of crucial biological entity interdependences including i.a. tumor response to treatment in the *generic investigational* context. To this end several combinations of mathematical concepts, entities and techniques have been developed and/or recruited and appropriately adapted. They include i.a. population dynamics models (Guiot et al 2006), cellular automata and hybrid techniques (Duechting and Vogelsaenger 1981; Duechting et al 1992; Ginsberg et al 1993; Kansal et al 2000; Stamatakos et al 2001a, 2001b; Zacharaki et al 2004), agent based techniques (Deisboeck et al 2001), diffusion related continuous and finite mathematics treatments (Murray 2003; Swanson et al 2002; Breward et al 2003; Cristini et al 2005; 1992; Frieboes et al 2006; Enderling et al 2007; Ramis-Conde et al 2008), etc. In addition, a number of large clinical tumor models focusing mainly on invasion and tumor growth morphology rather than on tumor response to concrete therapeutic schemes as administered in the clinical setting have appeared. Finite difference and finite element based solutions of the diffusion and classical mechanics equations constitute the core working tools of the corresponding techniques (Murray 2003; Swanson et al 2002; Clatz et al 2005).

Nevertheless, a number of concrete and pragmatic clinical questions of importance cannot be dealt with neither by the *bottom-up* approach nor by the morphologically oriented large tumor growth models in a direct and efficient way. Two examples of such questions are the following (Graf and Hoppe 2006): "Can the response of the local tumor and the metastases to a given treatment be predicted in size and shape over time?, What is the best treatment schedule for a patient regarding drugs, surgery, irradiation and their combination, dosage, time schedule and duration?" A promising modeling method designed with the primary aim of answering such questions is the *top-down* method developed by the *In Silico* Oncology Group (ISOG) ( *In Silico* Oncology Group , Stamatakos et al 2001c, 2002, 2006b, 2006c, 2006d, 2006e, 2007a, 2007b, 2009a; Dionysiou et al 2004, 2006a, 2006b, 2006c; 2007; 2008; Antipas et al 2004, 2007; Stamatakos and Uzunoglu 2006b; Stamatakos and Dionysiou 2009) . Macroscopic data, including i.a. anatomic and metabolic tomographic images of the tumor, provide the framework for the integration of *available* and *clinically trusted* biological information pertaining to lower and lower biocomplexity levels such as *clinically approved* histological and molecular markers. However, the method *does* also provide a powerful framework for the investigation of multiscale tumor biology in the *generic investigational* context.

From the mathematical point of view the *top-down* simulation method presented is primarily a *discrete* mathematics method, although *continuous* mathematics (continuous functions, differerential equations) are used in order to tackle specific aspects of the models such as pharmacokinetics and cell survival probabilities based on pharmacodynamical and radiobiological models. Adoption of the discrete approach as the core mathematical strategy of the method has been dictated by the obvious fact that from the cancer treatment perspective it is the *discrete* i.e. the integer number of the usually few biological cells surviving treatment and their discrete mitotic potential categorization (stem cells, progenitor cells of various mitotic potential levels and differentiated cells) that really matter. These *discrete* entities and quantities in conjunction with their complex interdependences may give rise to tumor relapse or to ensure tumor control over a given time interval following completion of the treatment course. Cell cycle phases have a clearly discrete character too. Moreover, the properties of the different cell phases may vary immensely from the clinical significance perspective. A classical example is the lack of effect of cell cycle specific drugs on living tumor cells residing in the resting G0 phase.

It is noted that complex interdependencies of microscopic factors in the surroundings of the cells such as oxygenation, nutrient supply and molecular signals emitted by other cells play a critical role in the mitotic fate of tumor cells. Their effect is taken into account in the method through the local mean values of the corresponding model parameters. To this end imaging, histological and molecular data is exploited as will be described further down.

Due to the numerical character of the method a careful and realistically thorough numerical analysis concerning consistency, convergence and sensitivity/stability issues is absolutely necessary before any application is envisaged. A discussion of this critical issue is included in Section 5.

Tumor neovascularisation is taken into account in an indirect yet pragmatic way by exploiting grey level and/or color information contained within slices of tomographic imaging modalities sensitive to blood perfusion and/or the metabolic status of the tumor. (Stamatakos et al 2001a,2002, 2006c; Dionysiou et al 2004, 2007; Marias et al 2007). The reason for adopting the above mentioned strategy rather than developing or integrating detailed tumor angiogenesis models is that no microscopic information regarding the exact mesh of the neovascularization *capillaries* throughout the tumor can be currently extracted from clinically utilized imaging modalities. Nevertheless, the microscopic functional capillary *density distribution* over the tumor can be grossly estimated based on various imaging modalities such as T1 gadolinium enhanced MRI in the case of glioblastoma multiforme (GBM) and arterial spin labelling (ASL) MRI.

Precursors of the method can be traced in the well established and clinically applicable disciplines of pharmacology and radiobiology. Integration of molecular biology in the *top-down* method may be viewed as the introduction of a perturbator or adaptor of the cellular and higher biocomplexity level parameters. In such a way *in vivo measurable* clinical manifestation of tumor dynamics is placed the foreground. This is one of the reasons why the method is gaining wider and wider acceptance within the clinical and the industrial environment including the emergent domain of *in silico* oncology (Stamatakos et al 2002, 2006a, 2007b, 2008, 2009a; Graf and Hoppe 2006; Graf et al 2007, 2009). Both the large scale European Commission (EC) and Japan funded ACGT research and development (R&D) project (ACGT) and the EC funded ContraCancrum R&D project (ContraCancrum) have adopted the *top-down* method as their core cancer simulation method. It is worth noting that in both projects the role of clinicians is prominent. A biomedical engineering concept and construct tightly associated with the method, the *Oncosimulator*, which is currently under clinical adaptation, optimization and validation is also sketched.

In order to convey the core philosophy of the method to the reader in a concise way a symbolic mathematical formulation of the *top-down* method in terms of a hypermatrix and discrete operators is introduced. Two specific models of tumor response to chemotherapeutic and radiotherapeutic schemes/schedules are briefly outlined so as to exemplify the method's application potential. The chapter concludes with a discussion of several critical aspects including numerical analysis, massive parallel code execution, associated technologies, extensions and validation within the framework of clinico-genomic trials and future challenges and perspectives.

A rather encouraging fact as far as industrial and eventually clinical translation of the method is concerned is that both the *top-down* method outlined and the *Oncosimulator* have been selected and endorsed by a worldwide leading medical technology company and now constitute modules of their research and development line (ContraCancrum). One of the envisaged final products of this endeavor is a radiotherapy treatment planning system based on both physical and multiscale biological optimization of the spatiotemporal dose administration scheme. A clinical trial based validation process for the system is currently at the final stage of its detailed formulation.

## 2. THE *ONCOSIMULATOR*

The *Oncosimulator* can be defined as a concept of multilevel integrative cancer biology, a complex algorithmic construct, a biomedical engineering system and (eventually in the future) a clinical tool which primarily aims at supporting the clinician in the process of optimizing cancer treatment in the patient individualized context through conducting experiments *in silico* i.e. on the computer. Additionally it is a platform for simulating, investigating, better understanding and exploring the *natural phenomenon* of cancer, supporting the design and interpretation of clinicogenomic trials and finally training doctors, researchers and interested patients alike (Stamatakos et Uzunoglu 2006b; Stamatakos et al 2007a; Graf et al 2009).

A synoptic outline of the clinical utilization of a specific version of the *Oncosimulator*, as envisaged to take place following an eventually successful completion of its clinical adaptation, optimization and validation process, is provided in the form of the following seven steps (Figure 1).

*1st step: Obtain patient's individual multiscale and inhomogeneous data.* Data sets to be collected for each patient include: clinical data (age, sex, weight etc.), eventual previous anti-tumor treatment history, imaging data (e.g. MRI, CT, PET etc images), histopathological data (e.g. detailed identification of the tumor type, grade and stage, histopathology slide images whenever biopsy is allowed and feasible etc.), molecular data (DNA array data, selected molecular marker values or statuses, serum markers etc.). It is noted that the last two data categories are extracted from biopsy material and/or body fluids.

*2nd step: Preprocess patient's data.* The data collected are preprocessed in order to take an adequate form allowing its introduction into the "Tumor and Normal Tissue Response Simulation" module of the *Oncosimulator*. For example the imaging data are segmented, interpolated, eventually fused and subsequently the anatomic entity/-ies of interest is/are three dimensionally reconstructed. This reconstruction will provide the framework for the integration of the rest of data and the execution of the simulation. In parallel the molecular data is processed via molecular interaction networks so as to perturb and individualize the average pharmacodynamic or radiobiological cell survival parameters.

*3rd step: Describe one or more candidate therapeutic scheme(s) and/or schedule(s).* The clinician describes a number of candidate therapeutic schemes and/or schedules and/or no treatment, obviously leading to free tumor growth, to be simulated *in silico* i.e. on the computer.

*4th step: Run the simulation.* The computer code of tumor growth and treatment response is massively executed on distributed grid or cluster computing resources so that several candidate treatment schemes and/or schedules are simulated for numerous combinations of possible tumor parameter values *in parallel* (see Section 5 for detailed justification). Predictions concerning the toxicological compatibility of each candidate treatment scheme are also produced.

*5th step: Visualize the predictions.* The expected reaction of the tumor as well as toxicologically relevant side effect estimates for all scenarios simulated are visualized using several techniques ranging from simple graph plotting to four dimensional virtual reality rendering.

*6th step: Evaluate the predictions and decide on the optimal scheme or schedule to be administered to the patient.* The Oncosimulator's predictions are carefully evaluated by the clinician by making use of their logic, medical education and even qualitative experience. If no serious discrepancies are detected, the predictions support the clinician in taking their final and expectedly optimal decision regarding the actual treatment to be administered to the patient.

*7th step: Apply the theoretically optimal therapeutic scheme or schedule and further optimize the Oncosimulator.* The expectedly optimal therapeutic scheme or schedule is administered to the patient. Subsequently, the predictions regarding the finally adopted and applied scheme or schedule are compared with the actual tumor course and a *negative feedback signal* is generated and used in order to optimize the *Oncosimulator*.

PLEASE PLACE FIGURE 1 (INCLUDING FIG. CAPTION) HERE

## 3. A BRIEF OUTLINE OF THE BASICS OF THE ISOG *TOP-DOWN* METHOD

### 3.1 The Multilevel Matrix of the Anatomical Region of Interest

The anatomical region of interest, primarily including the tumor and possibly adjacent normal tissues and edema, in conjunction with its biological, physical and chemical dynamics is represented by a multilevel matrix i.e. a

Matrix of (Matrices of (Matrices…of (Scalars or Vectors or Matrices )…)).

The multilevel matrix is created by a cubic discretization mesh which is virtually superimposed upon the anatomical region of interest. Biological cells residing within each geometrical cell of the mesh are conceptually clustered into mathematical *equivalence classes*. Equivalence classes primarily correspond to the various phases within or out of the cell cycle that a biological cell of the tumor can be found. Since a tumor cell at any given instant also belongs to a mitotic potential category (stem, progenitor, terminally_differentiated) the latter acts as a further partitioner of the biological cells into equivalence classes. One of the reasons, though *not* the single most important, for clustering biological cells into equivalence classes within each geometrical cell of the discretization mesh is computing resource limitations. Complex computational treatment of each single cell of a large clinical tumor undergoing therapeutic treatment as a *separate entity* is still not achievable within acceptable resource and time limits.

Discrete time is used. An important discretization aspect of the method is the mean time spent in the phase of an equivalence class by the biological cells belonging to the equivalence class (Stamatakos et

al 2001c, 2002, 2006c, 2006e; Dionysiou et al 2004, 2006; Stamatakos and Dionysiou 2009b). In order to allow for spatiotemporal perturbations of critical parameter values throughout the tumor and also avoid artificial cell synchronizations due to quantization, use of pseudo-random numbers is extensively made (Monte Carlo technique).

### 3.2. Practical Considerations Regarding the Construction of the Discretization Mesh

Collection of the appropriate mono-modality or far better multi-modality tomographic data of the patient such as slices of T1 weighted contrast enhanced MRI, T2 weighted MRI, CT, PET or other modalities, image segmentation, slice interpolation, three dimensional reconstruction of the anatomical entities of interest centered at the tumor, and eventually fusion of more than one modality images constitute the initial steps for the creation of the discretization mesh covering and discretizing the anatomical region of interest. Processed microscopic data (histological, molecular) are then utilized in order to enhance the patient individualization of the hypermatrix.

### 3.3 The basics of the top-down method

The multilevel matrix corresponding to the anatomical region of interest describes explicitly or implicitly the biological, physical and chemical dynamics of the region. The following parameters are used in order to identify a cluster of biological cells belonging to a given equivalence class within a geometrical cell of the mesh at a given time point:

I. the spatial coordinates of the discrete points of the discretization mesh with spatial indices i, j, k respectively. Each discrete spatial point lies at the center of a geometrical cell of the discretization mesh.

II. the temporal coordinate of the discrete time point with temporal index l

III. the mitotic potential category (i.e. stem or progenitor or terminally differentiated) of the biological cells with mitotic potential category index m

IV. the cell phase (within or out of the cell cycle) of the biological cells with cell phase index n. The following phases are considered: $\{G_1, S, G_2, M, G_0, A, N, D\}$ where $G_1$ denotes the $G_1$ cell cycle phase; S denotes the DNA synthesis phase; $G_2$ denotes the $G_2$ cell cycle phase; M denotes mitosis; $G_0$ denotes the quiescent (dormant) G0 phase; A denotes the apoptotic phase; N denotes the necrotic phase and D denotes the remnants of dead cells.

For the biological cells belonging to a given mitotic potential category *AND* residing in a given cell phase *AND* being accommodated within the geometrical cell whose center lies at a given spatial point *AND* being considered at a given time point - in other words for the biological cells clustered in the same equivalence classs denoted by the index combination ijklmn - the following state parameters are provided

i. local oxygen and nutrient provision level. The following possible binary values of this parameter were initially considered: “oxygen and nutrient provision level *sufficient* for tumor cell proliferation” and “oxygen and nutrient provision level *insufficient* for tumor cell proliferation”. Obviously the binary character of the oxygen and nutrient provision level is to be considered only a first simplifying approximation. More elaborate descriptions have been proposed and applied (Stamatakos et al 2002, 2006c, 2006d; Dionysiou et al 2004, 2006a; Antipas et al 2004).

ii. number of biological cells

iii. average time spent by the biological cells in the given phase

iv. number of biological cells hit by treatment

v. number of biological cells *not* hit by treatment

The initial constitution of the tumor i.e. its biological, physical and chemical state has to be estimated based on the available medical data through the application of pertinent algorithms ( Kolokotroni et al 2008; Georgiadi et al 2008) . This state corresponds to the instant just before the start of the treatment course to be simulated.

The entire simulation can be viewed as the periodic and sequential application of a number of sets of algorithms on the multilevel matrix of the anatomical region of interest. Thus a stepwise multilevel matrix updating is achieved. Each *algorithm set application period* is equal to the time separating two consecutive *complete scans* of the discretization mesh. A complete scan includes mesh scans performed by all algorithm sets for any given time point. The algorithm set application period is usually taken 1 h since this is approximately the duration of mitosis, the shortest of the cell cycle phases. It should be noted that although the parameter values *exported* by the simulation execution at any desired instant for visualization and analysis purposes have a discrete character, certain parameters are handled by the computer *internally* and *temporarily* as real numbers (even with enhanced precision) in order to minimize quantization error propagation, in particular when dealing with small numbers of discrete entities in the stochastic context. By no means, however, does this technicality affect the fundamentally discrete character of the *top-down* method described.

The application of the algorithm sets on the multilevel matrix of the anatomic region of interest takes place in the following order:

A. Time updating i.e. increasing time by a time unit (e.g. 1h)

B. Estimation of the local oxygen and nutrient provision level

C. Estimation of the effect of treatment (therapy) referring mainly to cell hitting by treatment, cell killing and cell survival.

D. Application of cell cycling, possibly perturbed by treatment. Transition between mitotic potential cell categories such as transition of the offspring of a terminally divided progenitor cell into the terminally differentiated cell category is also tackled by this algorithm set.

E. Differential expansion or shrinkage or more generally geometry and mechanics handling.

F. Updating the local oxygen and nutrient provision level following application of the rest of algorithm sets at each time step

It is noted that the outcome of appropriate processing of the molecular and/or histopathological data via e.g. molecular networks and signaling pathways is used as a perturbator of the cell survival probabilities included in algorithm set "C" so as to considerably enhance patient individualization of the simulation. A realistic estimate of the extent of such perturbations for a given tumor type subclass in the framework of a clinico-genomic trial is achieved in a stepwise way. Initial rough modifications of the cell survival probabilities based on the baseline-pretreatment data, pertinent literature information and logic are subsequently corrected through utilization of the corresponding post treatment data via a process of parameter fitting.

Obviously the above mentioned concepts and briefly outlined steps cannot convey all the details needed for the simulation to run. Their role is rather to identify and decompose the major conceptual mathematical and computational steps than to list all modeling details. The interested reader is referred to the website of the *In Silico* Oncology Group where they may find lists of pertinent publications providing detailed descriptions of several *top-down* multilevel cancer models including i.a. assumptions, mathematical treatment, numerical aspects such as convergence and quantization error minimization, sensitivity analysis, validation, applications and suggested extensions.

It is worth noting that discrete simulation under certain constraints can efficiently replace analytical solutions to a wide range of mathematical problems which, although being formulated in terms of *continuous* mathematics - usually including symbolically formulated differential equations - refer in fact to *discrete* physical quantities such as biological cells and cell state transition rates. Moreover, in many cases the *continuous* symbolic formulation of mathematical operators, such as the well known differential operator, when acting on discrete physical quantities can be readily replaced by a conceptually more straightforward *algorithmic* formulation. Several techniques leading to the minimization of error propagation for those cases where small numbers of discrete entities are dealt with by stochastic processes are available. The above generic policy has been extensively adopted in top-down multiscale models. For extensions of the method currently under implementation see Section 5.

## 4. EXAMPLES OF APPLICATIONS OF THE ISOG *TOP-DOWN* METHOD - RESULTS

In this section two indicative models denoted by Model A and Model B are briefly outlined so as to exemplify the application potential of the method in the clinical context.

### 4.1 Model A: Tumor Response to Chemotherapeutic Schedules

Model A is a four dimensional, patient specific *top-down* simulation model of solid tumor response to chemotherapeutic treatment *in vivo*. The special case of imageable glioblastoma multiforme (GBM) treated by temozolomide (TMZ) has been addressed as a simulation paradigm. However, a considerable number of the involved algorithms are quite generic. The model is based on the patient's imaging, histopathologic and genetic data. For a given drug administration schedule lying within acceptable toxicity boundaries, the concentration of the prodrug and its metabolites within the tumor is calculated as a function of time based on the drug pharamacokinetics. A discretization mesh is superimposed upon the anatomical region of interest and within each geometrical cell of the mesh the basic biological, physical and chemical *"laws"* such as the rules concerning oxygen and nutrient provision, cell cycling (Salmon and Sartorelli 2001), mechanical deformation etc. are applied at each discrete time point. The biological cell fates are predicted based on the drug pharmacodynamics (Newlands et al 1992; Bobola et al 1996; Perry 2001; Katzung 2001; FDA; Chinot et al 2001; Stupp et al 2001). The outcome of the simulation is a prediction of the spatiotemporal activity of the entire tumor and is i.a. virtual reality visualized. A good qualitative agreement of the model's predictions with clinical experience (Stamatakos et al 2006d, 2006e) supports the applicability of the approach. Model A has provided a basic platform for performing patient individualized *in silico* experiments as a means of chemotherapeutic treatment optimization in the *theoretical* context. A few indicative aspects of the model are outlined below. Since the complexity of the analysis is high the interested reader is referred to (Stamatakos et al 2006d, 2006e) for a detailed description of the model. The work has also provided the basis for the development of the ACGT (ACGT) and ContraCancrum (ContraCancrum) chemotherapy treatment response models.

**PLEASE PLACE FIGURE 2 (INCLUDING FIGURE CAPTION) HERE**

Figure 2 depicts the simplified cytokinetic model of a tumor cell that has been proposed and adopted in Model A. The cytotoxicity produced by TMZ is primarily modeled by a delay in the S phase compartment (TDS) which is denoted by "Delay due to the effect of chemotherapy" in the diagram of Figure 2 and by subsequent apoptosis. Further details are provided in the caption of Figure 2.

**PLEASE PLACE FIGURE 3 (INCLUDING FIGURE CAPTION) HERE**

Figure 3 provides a three dimensional visualization of the simulated response of a clinical GBM tumor to one cycle of the TMZ chemotherapeutic scheme: 150 mg/m$^2$ orally once daily for 5 consecutive days per 28-day treatment cycle (Stamatakos et al 200d). Panel (a) shows the external surface of the tumor before the beginning of chemotherapy. Panel (b) shows the internal structure of the tumor before the beginning of chemotherapy. Panel (c) shows the predicted external surface of the tumor 20 days after the beginning of chemotherapy. Panel (d) shows the predicted internal structure of the tumor 20 days after the beginning of chemotherapy. The pseudo-coloring criterion proposed and utilized is described in the caption of Figure 3.

### 4.2 Model B: Tumor Response to Radiotherapeutic Schedules

Model B is a spatiotemporal simulation model of *in vivo* tumor growth and response to radiotherapy exemplified by the special case of imageable GBM treated by the treatment modality under consideration. The main constitutive processes of the model can be summarized as follows. A discretizing cubic mesh is superimposed upon a three dimensional virtual reconstruction of the tumor including its necrotic region and the surrounding anatomical features based on imaging data. In a way analogous to Model A, within each geometrical cell of the mesh a number of biological cell equivalence classes are defined based i.a. on the biological cell distribution over the various phases within or out of the cell cycle for the various mitotic potential categories. Sufficient registers are used in order to store the current state of each equivalence class such as the average time spent by clustered biological cells in phase G1 etc. The mesh is scanned every one hour. The basic *biological, physical* and *chemical "laws"* or more precisely rules including the metabolic activity dynamics, cell cycling, mechanical and geometrical aspects, cell survival probability following irradiation with dose D (Perez and Brady 1998; Steel 2002) are applied on each geometrical cell at each complete scan. A spatial and functional restructuring of the tumor takes place during each discrete time point since new biological cells are eventually produced, leading to differential tumor growth, or existing cells eventually die and subsequently disappear through specific molecular and cellular event cascades, thus leading to differential tumor shrinkage. Simulation predictions can be two or three dimensionally visualized at any simulated instant of interest. In the particular model special attention has been paid to the influence of oxygenation on radiosensitivity in conjunction with the introduction of a refined imaging based description of the neovasculature density distribution. In order to validate the model two identical - except for the p53 gene status - virtual GMB tumors of large size, complex shape and complex internal necrotic region geometry were considered. The first one possessed a wild type p53 gene whereas the second one was characterized by a mutated p53 (Stamatakos et al 2006c; Dionysiou et al 2004). The values of the α and β parameters of the standard linear quadratic radiobiological model for cell survival (Steel 2002, Perez and Brady 1998) have been determined experimentally for the two cell lines considered (Haas-Kogan et al 1995). Simulation predictions agree at least semi-quantitatively with clinical experience and in particular with the outcome of the Radiation Therapy Oncology Group RTOG Study 83 02 (Werner-Wasik et al 1996). The model allows for a quantitative study of the inter-relationship between the competing influences in a complex, dynamic tumor environment. Therefore, the model is already useful as an educational tool with which to theoretically study, understand and demonstrate the role of various parameters on tumor growth and response to irradiation. A long term quantitative clinical adaptation and validation of a considerably extended version of the model is in progress within the framework of the ContraCancrum project (ContraCancrum). The long term goal is obviously integration into the clinical treatment planning procedure.

Figure 4 shows simulation predictions corresponding to several branches of the RTOG Study 83 02. Panel (a) provides the total number of proliferating and dormant tumor cells as a function of time for the hyperfractionated (1.2 Gy twice daily, 5 days per week to the dose of 72 Gy, “HF-72”) and accelerated hyperfractionated (1.6 Gy twice daily, 5 days per week to the dose of 48 Gy, “AHF-48”) radiotherapy schedules. All schemes start on the first day of the radiotherapy course. HF-72 is completed on day 40 after initiation of treatment whereas AHF-48 is completed on day 19. Figure 4(b) shows the total number of proliferating and dormant tumor cells as a function of time for the hyperfractionated (1.2 Gy twice daily, 5 days per week to the dose of 64.8 Gy, “HF-64.8”) and accelerated hyperfractionated (1.6 Gy twice daily, 5 days per week to the dose of 48 Gy, “AHF-48”) radiotherapy schedules. Both irradiation schedules start on the first day of the first week of treatment. Both irradiation schedules start on the first day of the first week of treatment. HF-64.8 is completed on day 37 after initiation of treatment whereas AHF-48 is completed on day 19. According to the graphs, before completion of the AHF course cell kill due to AHF irradiation is more pronounced than cell kill induced by the HF scheme. This can be explained by the fact that a higher total dose has been administered to the tumor by the AHF scheme whereas for the period under consideration both schemes are characterized by the same time intervals between consecutive sessions. In case that not all living cells have been killed by AHF irradiation, tumor repopulation is considerable so that by the time the HF scheme is completed living tumor cells and their progeny which have escaped AHF irradiation outnumber tumor cells which have escaped HF irradiation. Improved tumor control following HF irradiation in comparison with the AHF scheme is in agreement with the conclusions of the clinical trial RTOG-83-02.

**PLEASE PLACE FIGURE 4 (INCLUDING FIGURE CAPTIONS) HERE**

## 5. DISCUSSION

Since the *top-down* multilevel method presented is a numerical method, a thorough convergence and sensitivity/stability analysis that includes the study of multiple parameter interdependences is necessary before any application is envisaged. Numerical analysis should satisfactorily cover at least those regions of the abstract parameter space that correspond to the envisaged applications. It is noted that of particular importance is the creation of the baseline tumor constitution by exploiting the relevant multilevel data available. Convergence of the tumor initialization has to be ensured. All of the above issues have been successfully addressed for specific tumor treatment cases such as breast cancer treated with epirubicin and nephroblastoma treated with vincristine and dactinomycin. The numerical behavior of the corresponding models has been checked through massive numerical experimentation. Concrete applicability intervals, restrictions and limitations have been identified (Internal ACGT project reports and deliverables; Kolokotroni et al 2008; Georgiadi 2008). Since the entire parameter space of the *top-down* models is rather large, numerical studies covering regions that correspond to further applications are in progress. Special attention is paid to the inherent relative biological instability of the cancer system itself when the model's stability is investigated.

It is well known that the values of critical parameters determining treatment outcome can vary considerably around what is assumed to be their population based average values. *Even after* incorporation of patient specific multiscale data into the simulation model no *accurate* evaluation of several critical model parameters is expected to be achieved. Moreover, as already mentioned, a tumor may behave as a relatively unstable system. Therefore, in order to compare candidate treatment schemes and/or schedules *in silico*, several possible combinations of parameter values lying around their apparently most probable estimates have to be constructed so as to cover the abstract parameter space as best as possible. Code executions have to be performed for all these selected parameter combinations. If for example the clinical question addressed is "Which out of the two candidate treatment schedules dented by I and II is the most promising for a given patient?", simulations have to run for both schedules I and II and for all parameter value combinations selected in the way briefly delineated above. If based on the simulation predictions schedule I outperforms schedule II for a sufficiently large percentage of the total parameter combinations considered, say 90%, then there is ground to suggest adoption of schedule I. Candidate scheme/schedule selection criteria are currently under formulation in tight collaboration with specialist clinicians within the framework of the ACGT (ACGT) and ContraCancrum (ContraCancrum) R&D projects. Obviously the above sketched treatment optimization strategy dictates the need for a large number of parallel code executions on either cluster of grid platforms. This necessity has been addressed by specific actions of the previosly mentioned projects.

Critical constraints imposed by toxicological limits of the treatment affected normal tissues should also be taken into account in order to judge whether or not a candidate scheme could be toxicologically acceptable. This issue may be addressed by exploiting the outcome of eventually relevant clinical trials and in particular of their phase I results. Ideally, direct multiscale spatiotemporal simulation of the effects of a given candidate scheme on specific normal tissues would provide quantitatively refined predictions. Nevertheless, due to the extremely high complexity of the homeostatic mechanisms governing normal tissue dynamics, the large number of normal tissue functional aspects and the potential induction of serious late effects by treatment such as radiotherapy, clinical translation of the second scenario seems to be a long term enterprise (Antipas et al 2007).

Since many solid tumors are microscopically inhomogeneous in space, the applications presented so far essentially make use of the *mean values* of certain biological parameters over each *imaging based segmented* sub-region of the tumor ( Stamatakos et al 2002, 2006b, 2006c, 2006d, 2006e, 2007b, Dionysiou et al 2004, 2006a, 2006c; 2007; 2008; Antipas et al 2004, 2007). Small perturbations around these values are nevertheless implemented across each region through Monte Carlo simulation by the *top-down* method presented. In the paradigmal case of MRI T1 gadolinium enhanced imaging modality, strong grey level fluctuations over a tomographic slice can lead to an approximate delineation of the internal necrotic and the well neovascularized region of the tumor. Despite the fact that different values of certain parameters may be assigned to these two regions, sub-imaging scale inhomogeneties may still create spatial fluctuations of certain parameter values. In order to theoretically investigate the role of such biological inhomogeneities, pertaining for example to the genotypic and/or phenotypic tumor constitution, as well as the role of biochemical inhomogeneities of the extra tumoral environment such as acidity, necrosis exudate concentration etc the *top-down* basic platform can be still used provided that specific adaptations have taken place. Furthermore, tumor cell - tumor cell, tumor cell - host cell and tumor cell - local environment interactions in the microscopic setting can in principle be studied. In order to implement the above scenarios, the density of the discretization mesh should considerably increase, deeper level partitioning into more equivalence (sub-)classes has to be introduced into the multilevel matrix of the anatomical region of interest and the algorithm sets should be extended accordingly. However, such an approach dictates a sharp increase in computing memory and time demands and therefore tumor size must be kept small if restrictions in these resources apply, as is usually the case.

Following appropriate adaptation of specific modeling modules or equivalently algorithm sets such as the “C” set referring to cell killing etc. the *top-down* method outlined is in principle able to simulate tumors of any shape, size, geometry, macroscopic distribution of the metabolic or neovascularisation field, differentiation grade, spatial inhomogeneities, molecular profile and treatment scheme/schedule such as radiotherapeutic, chemotherapeutic, combined, new treatment modality etc. However, great care should be taken so that the model parameter values are estimated as best as possible based on real multilevel data. If such data is not available use of at least population based average parameter value estimates or qualitative experience based plausible values may be utilized only for *generic exploratory* reasons.

Hybridization of the *top-down* method presented with continuous and finite mathematics approaches such as diffusion based tumor growth modeling and detailed biomechanics is currently under implementation (ContraCancrum). The aim of the task is to integrate into a *top-down* GBM model the microscopic tumor invasion process. The detailed biomechanics of the system calculated via a finite element module is also being integrated. Such a hybrid model is expected to be able to reproduce in relative detail both *physical and biological* aspects of tumor dynamics within the *generic investigational* framework. It should be noted that the non imageable diffusive component of GBM does play an important role in the development of the disease and therefore merits an in depth theoretical investigation. However, since the non imageable boundaries of GBM cannot be defined and monitored in a sufficiently objective way i.e. based on observational data such as clinically obtainable tomograhic images, direct handling of the non imageable component by treatment planning systems in the patient individualized treatment context seems not to be a fully mature scenario as yet. Furthermore, by focusing on the imageable component within the treatment optimization context, one may argue that if for the imageable component a candidate treatment scheme denoted by scheme I outperforms another candidate scheme denoted by scheme II *in silico*, the same would be true for the non imageable component of the tumor as well. From the *treatment perspective* again the main advantage of focusing on the imageable component, although this may represent even less than half of the total number of all viable tumor cells, is that this very component is amenable to relatively objective measurement *in vivo* and not only *post mortem*. Therefore, glioma dynamics models based on the imageable component are amenable to validation, at least in part, *in vivo.* Besides, incorporation of the immune system response to tumor (D'Onofrio 2005) and simulation of the effects of antiangiogenetic drugs on the tumor are two further scenarios, currently under investigation in the ACGT Oncosimulator extension context (ACGT).

Referring to the molecular level from the *generic investigational* standpoint, a large number of mechanisms, such as pathways leading to apoptosis or survival, that can be informed by available molecular data can be readily integrated into *top-down* models by applying the *summarize and jump strategy* of bio-data and bio-knowledge integration across bio-complexity scales (Stamatakos et al 2009a). This is in fact one of the actions currently taking place within the framework of the ContraCancrum project (ContraCancrum). However, if the same biocomplexity level is viewed from the *clinical* perspective, care has to be taken so that only those characteristics and /or mechanisms whose predictive potential has been proved and established in the *clinical* setting - normally through clinical trials – may be incorporated into the models.

Regarding the envisaged clinical translation of *top-down* based models and systems, including the *Oncosimulator*, a *sine qua non* prerequisite is a systematic, formal and strict clinical validation. Designing the models so as to mimick actual clinical or far better *clinico-genomic* trials seems to be the optimal way to achieve this goal (Stamatakos et al. 2006c, 2007a; Graf and Hoppe 2006; Graf 2007, 2008, 2009). Therefore, involvement of clinicians in the model and system design and validation process should start at the *very beginning* of the endeavor (Graf 2007, 2008, 2009). Real clinicogenomic trials can provide invaluable multiscale data (imaging, histological, molecular, clinical, treatment) before, during and after a treatment course so as to best adapt and optimize the models and subsequently validate them. This is one of the core tasks of both the ACGT (ACGT) and the ContraCancrum R&D (ContraCancrum) projects. Nephroblastoma and breast cancer are the tumor types addressed by ACGT whereas gliomas and lung cancer are the ones addressed by ContraCancrum.

A further important challenge is to develop reliable, efficient, highly versatile and user friendly technological platforms which, following clinical adaptation, optimization and validation of the models would facilitate translation of *oncosimulators* into the clinical practice so as to efficiently support, enhance and accelerate patient individualized treatment optimization. Advanced image processing, visualization and parallel code execution modules are but a few of the components necessary to achieve this goal. Both the ACGT and ContraCancrum R&D projects constitute exemplary initiatives towards this direction.

In summary both the *top-down* multilevel cancer simulation method briefly outlined above and the *Oncosimulator* have been designed so as to be readily optimizable, extensible and adaptable to changing clinical, biological, and research envirionments. Thus both entities, being primarily *multiscale physics* and *biomedical engineering* geared, have a *pragmatic* and *evolutionary* character.

## ACKNOWLEDGEMENTS

This work has been supported in part by the European Commission under the projects "ACGT: Advancing Clinicogenomic Trials on Cancer" (FP6-2005-IST-026996) and ContraCancrum: Clinically Oriented Translational Cancer Multilevel Modelling" (FP7-ICT-2007-2- 223979).

## CAPTIONS

**Figure 1:** The *Oncosimulator*: a gross workflow diagram

**Figure 2:** Simplified cytokinetic model of a tumor cell proposed and adopted in model A. Symbols: G1: G1 phase; S: DNA synthesis phase; G2: G2 phase; G0: G0 phase; N: necrosis; A: apoptosis. The cytotoxicity produced by TMZ is primarily modeled by a delay in the S phase compartment (TDS) ("Delay due to the effect of chemotherapy" in the diagram) and subsequent apoptosis. The delay box simply represents the time corresponding to at most two cell divisions being required before the emergence of temozolomide cytotoxicity. It is not a time interval additional to the times represented by the cell cycle phase boxes. (From Stamatakos G. S., Antipas V. P. and N.K.Uzunoglu. 2006. A spatiotemporal, patient individualized simulation model of solid tumor response to chemotherapy *in vivo*: the paradigm of glioblastoma multiforme treated by temozolomide. IEEE Trans Biomed Eng 53(8): 1467- 1477. Reprinted with permission from the Institute of Electrical and Electronics Engineers (IEEE), © 2006 IEEE)

**Figure 3:** A three dimensional visualization of the simulated response of a clinical GBM tumor to one cycle of the TMZ chemotherapeutic scheme: 150 $mg/m^2$ orally once daily for 5 consecutive days per 28-day treatment cycle. (a) external surface of the tumor before the beginning of chemotherapy, (b) internal structure of the tumor before the beginning of chemotherapy, (c) predicted external surface of the tumor 20 days after the beginning of chemotherapy (d) predicted internal structure of the tumor 20 days after the beginning of chemotherapy. The following pseudocolor code has been applied: *red*: proliferating cell layer; *green*: dormant cell layer (G0); *blue*: dead cell layer. The "99.8%" pseudo-coloring criterion has been devised and applied as follows. "For a geometrical cell of the discretizing mesh, if the percentage of dead cells within it is lower than 99.8% then {if percentage of proliferating cells > percentage of G0 cells, then paint the geometrical cell red (proliferating cell layer), else paint the geometrical cell green (G0 cell layer)} else paint the geometrical cell blue (dead cell layer).” (From Stamatakos G. S., Antipas V. P. and N.K.Uzunoglu. 2006. A spatiotemporal, patient individualized simulation model of solid tumor response to chemotherapy *in vivo*: the paradigm of glioblastoma multiforme treated by temozolomide. IEEE Trans Biomed Eng 53(8): 1467- 1477. Reprinted with permission from the Institute of Electrical and Electronics Engineers (IEEE), © 2006 IEEE)

**Figure 4:** Simulation predictions corresponding to several branches of the RTOG Study 83 02. (a) Total number of proliferating and dormant tumour cells as a function of time for the hyperfractionated (1.2 Gy twice daily, 5 days per week to the dose of 72 Gy, “HF-72”) and accelerated hyperfractionated (1.6 Gy twice daily, 5 days per week to the dose of 48 Gy, “AHF-48”) radiotherapy schedules. HF-72 is completed on day 40 after initiation of treatment whereas AHF-48 is completed on day 19. (b) Total number of proliferating and dormant tumour cells as a function of time for the hyperfractionated (1.2 Gy twice daily, 5 days per week to the dose of 64.8 Gy, “HF-64.8”) and accelerated hyperfractionated (1.6 Gy twice daily, 5 days per week to the dose of 48 Gy, “AHF-48”) radiotherapy schedules. HF-64.8 is completed on day 37 after initiation of treatment whereas AHF-48 is completed on day 19. Both irradiation schedules start on the first day of the first week of treatment. (From Stamatakos G.S., V.P. Antipas, N.K. Uzunoglu, and R.G. Dale. 2006. A four dimensional computer simulation model of the *in vivo* response to radiotherapy of glioblastoma multiforme: studies on the effect of clonogenic cell density. Br J Radiol 79: 389-400. Reprinted with permission from the British Institute of Radiology, © 2006 BIR)

# FIGURES

## FIGURE 1.

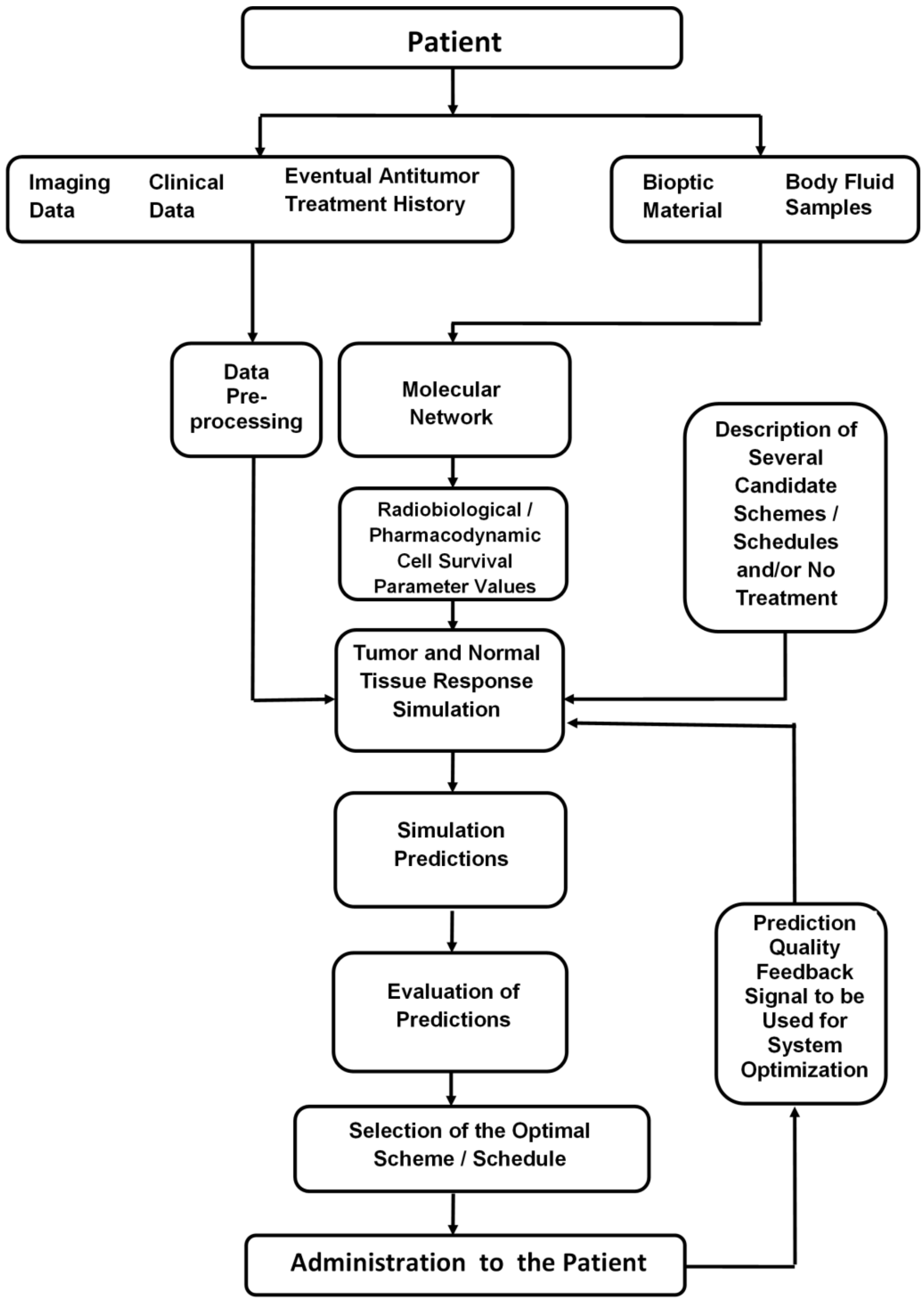

**FIGURE 2.**

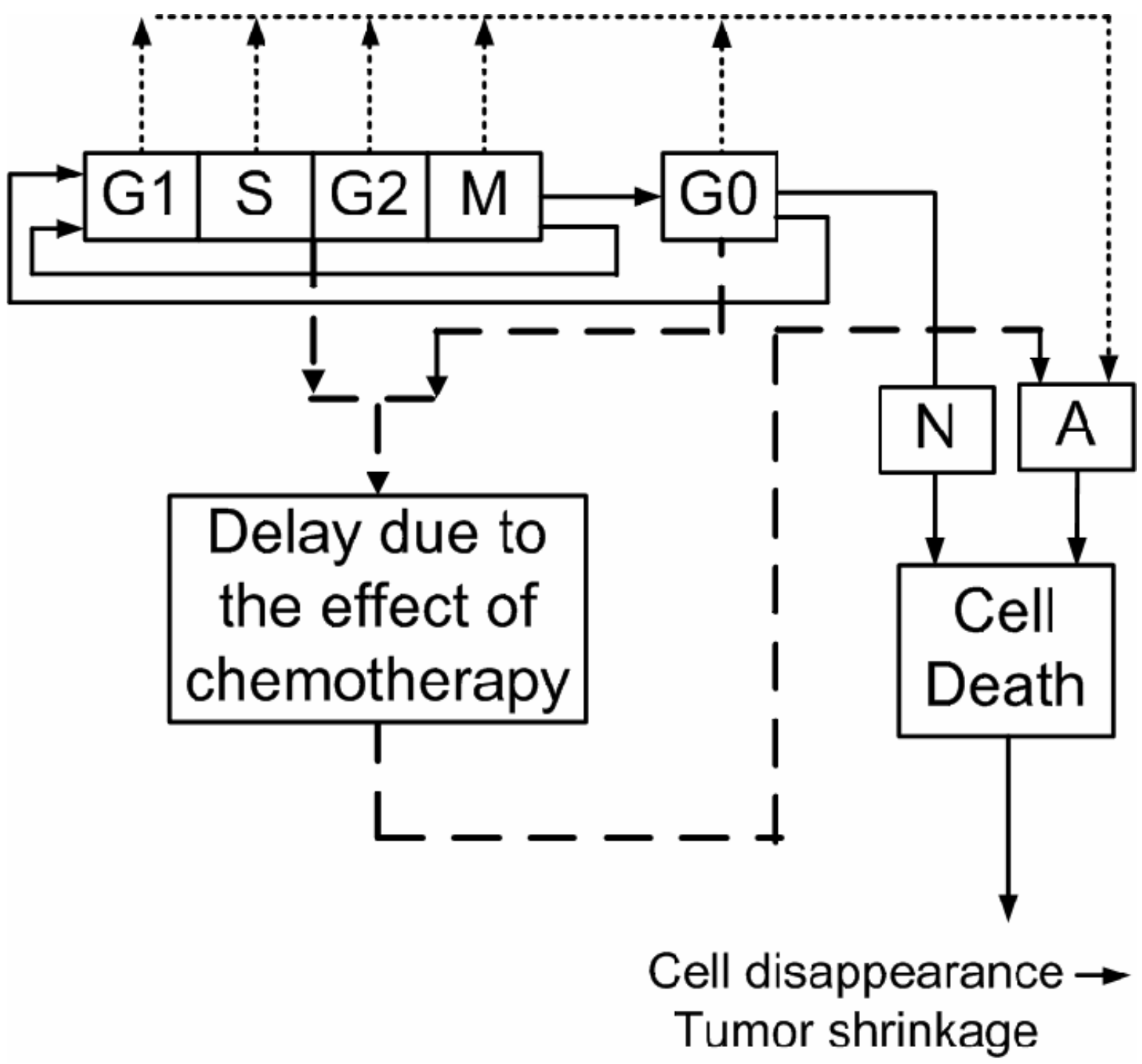

**FIGURE 3.**

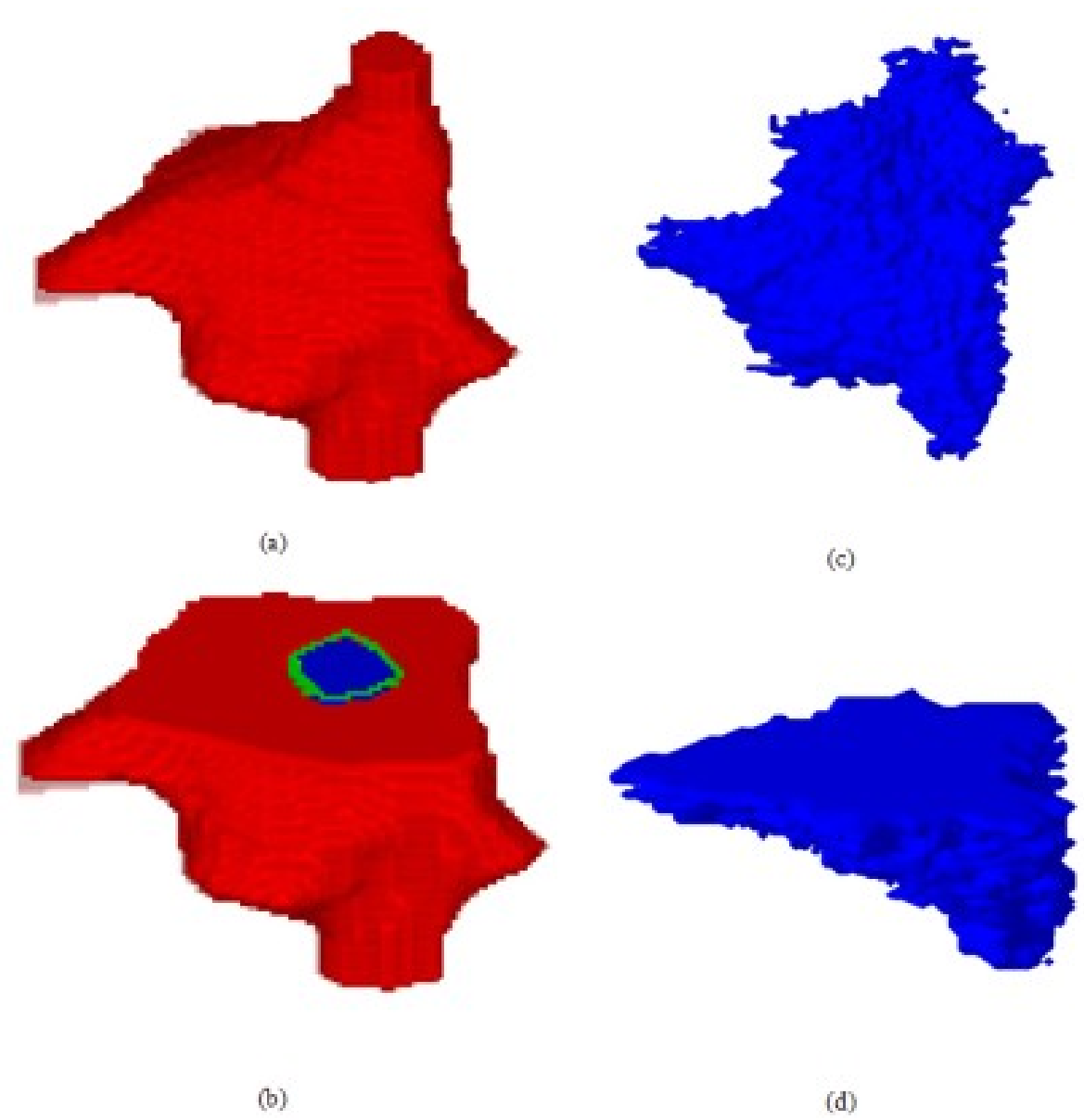

**FIGURE 4.**

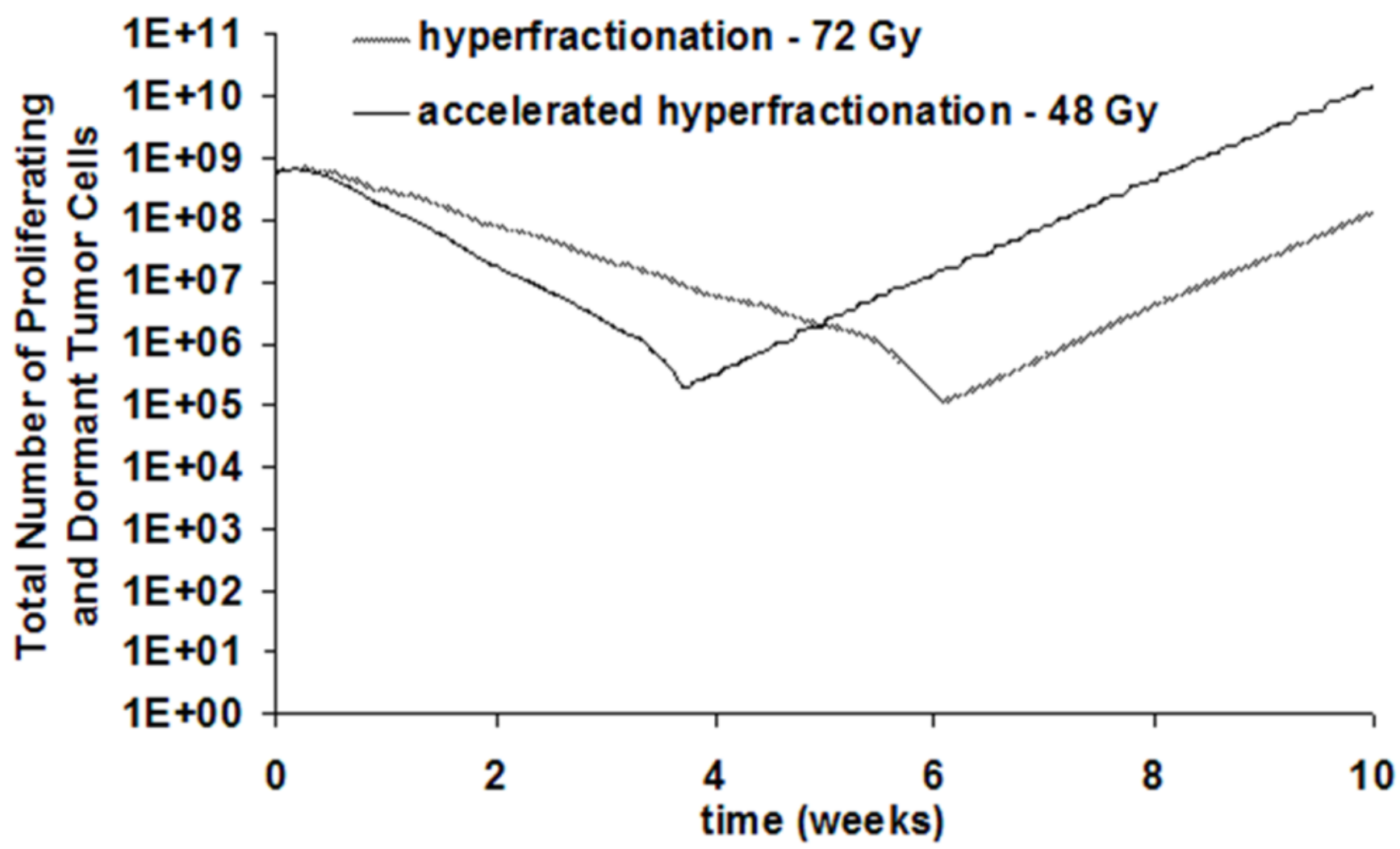
hyperfractionation - 72 Gy
accelerated hyperfractionation - 48 Gy
Total Number of Proliferating and Dormant Tumor Cells
1E+11
1E+10
1E+09
1E+08
1E+07
1E+06
1E+05
1E+04
1E+03
1E+02
1E+01
1E+00
0
2
4
6
8
10
time (weeks)

(a)

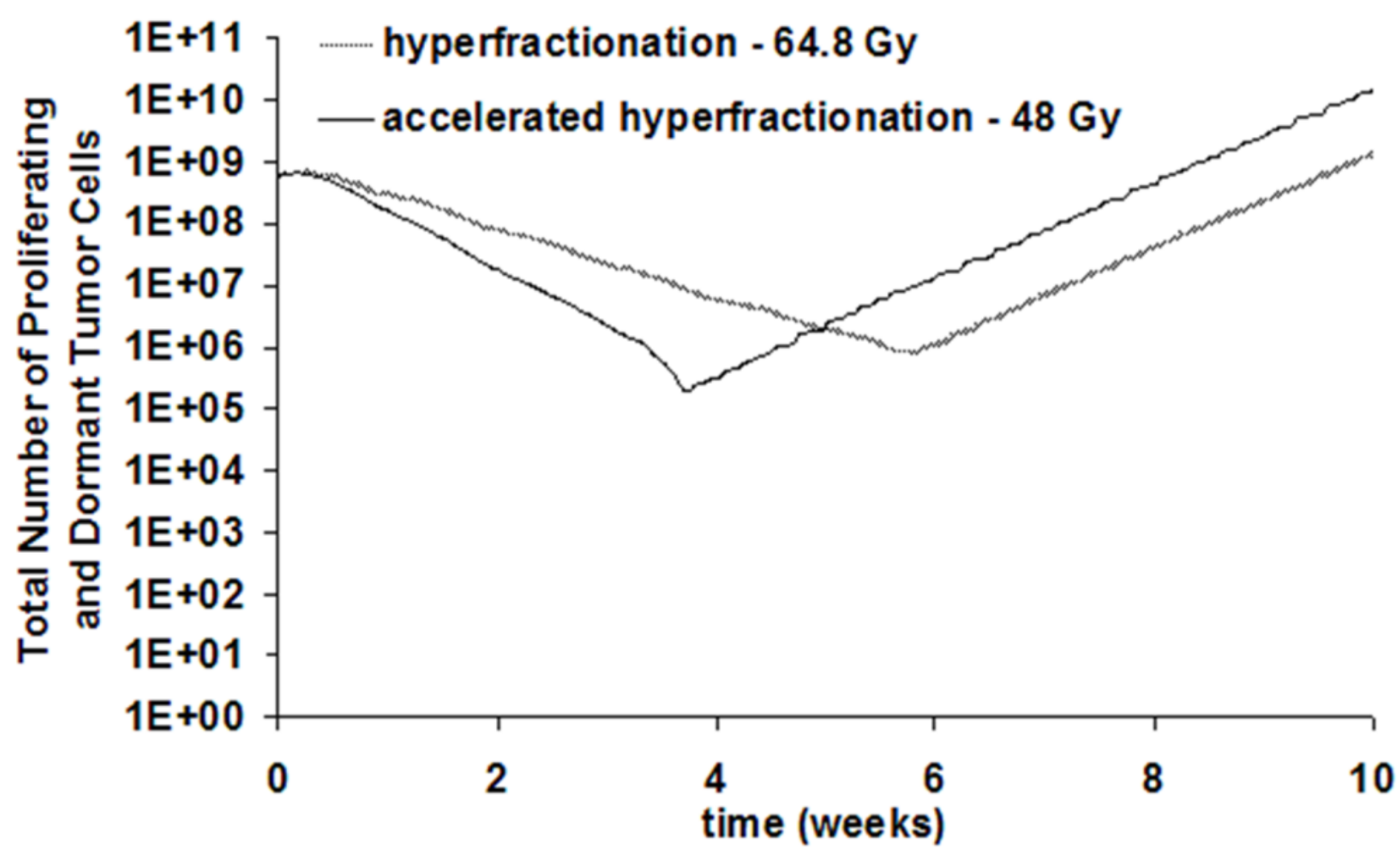
hyperfractionation - 64.8 Gy
accelerated hyperfractionation - 48 Gy
Total Number of Proliferating and Dormant Tumor Cells
1E+11
1E+10
1E+09
1E+08
1E+07
1E+06
1E+05
1E+04
1E+03
1E+02
1E+01
1E+00
0
2
4
6
8
10
time (weeks)

(b)